\documentclass[aps,reprint,prl,twocolumn,superscriptaddress]{revtex4-1}
\usepackage{color}
\usepackage{graphicx}

\begin{document}

\title{Novel Differential Conductance Oscillations in Asymmetric Quantum Point Contacts}
\author{Hao Zhang}
\affiliation{Department of Physics, Duke University, Physics Building, Science Drive, Durham, North Carolina 27708, USA}
\author{Phillip M. Wu}
\affiliation{Department of Applied Physics and Geballe Laboratory for Advanced Materials, Stanford University, Stanford, California 94305}
\author{A. M. Chang}
\affiliation{Department of Physics, Duke University, Physics Building, Science Drive, Durham, North Carolina 27708, USA}

\date{\today}

\begin{abstract}

Small differential conductance oscillations as a function of source-drain bias were observed and systematically studied in an asymmetric quantum point contact (QPC). These oscillations become significantly suppressed in a small in-plane magnetic field ($\sim~0.7~T$) or at higher temperatures ($\sim~800~mK$).  Qualitatively, their temperature evolution can be simulated numerically based on smearing of the Fermi distribution, whereas features near zero-bias cannot. Single particle scenarios are unsatisfactory in accounting for the oscillations, suggesting that they are likely caused by electron and spin correlation effects.

\end{abstract}

\maketitle

After years of search, the elusive quantum Wigner crystal-like state in 1-dimension and quasi-1-dimension, a state in which an uncertainty in the momentum yields a relevant kinetic energy scale, but Coulomb repulsion wins out--still has not been unequivocally demonstrated. The most promising system, in which a systematic study can be carried out using transport measurements, is the QPC system. Despite the lack of a smoking-gun signature, such as a threshold for conduction analogous to the motion of a pinned charge density wave (or at least a highly nonlinear increase beyond some pinning voltages), to date, several puzzling characteristics have been discovered in ballistic QPCs, when the number of conduction channels (transverse modes) is tuned toward the single channel limit. These include the well-studied 0.7 anomaly \cite{Thomas,Micolich}, the suppression (or destruction) of the first quantized conductance plateau, \cite{HewPepper1,HewPepper2,Wu,Hao} as well as higher plateaus \cite{HewPepper2,Hao_thesis}, and the recently discovered conductance resonances in geometrically asymmetric QPCs \cite{Wu,Hao}.  None of these features can be explained adequately without introducing Coulomb interaction and spin effects.

In this letter, we report the discovery of a new signature unique to the QPC system.  Specifically, we focus on unusual oscillations in the differential conductance ($dI/dV$) as a function of source-drain bias, in an asymmetric ballistic QPC. The magnetic field behavior shows that these oscillations are readily suppressed under a small in-plane magnetic field ($\sim~0.7T$), while the temperature dependence shows that these oscillations are washed out at around $T=800~mK$. This type of oscillations as a function of source-drain bias has not been reported in the published literature in any electronic system \cite{Liu2009Brun2014}; the only possible exception is unpublished work in a quadruple-quantum-dot, which is entirely different from the QPC system \cite{Shang2013}.  Similar oscillations are reproduced is a second sample at $300~mK$ and zero magnetic field. It is noteworthy that in the quadruple quantum dot (QD) system--arranged in two rows of double-QDs, the oscillations, at least two (and up to four) to each side of zero bias, were likely of a Kondo origin.  This raises the possibility that in our QPC system, Kondo physics is involved, which arises when the electron system spontaneously forms a double-row of localized puddles, suggestive of a double row lattice.

Our QPC samples were fabricated by electron beam lithography, evaporation of the Cr/Au surface gates and lift-off. The two-dimensional electron gas (2DEG) is located at a shallow depth 85 nm below the surface of the GaAs/AlGaAs heterostructure crystal. The carrier density and mobility are $3.8\times 10^{11} /cm^2$ and $9\times 10^5 cm^2/Vs$ respectively, giving a mean free path $\sim 9~\mu m$.  An excitation voltage $V_{ac}=5~\mu V$ at 17.3 Hz was applied, and the current was measured in a PAR124A lock-in amplifier after conversion to a voltage using a home-made current preamplifier. The measurement was first carried out in a He-3 fridge with a base temperature $300mK$, and then repeated in a dilution refrigerator with a base temperature below $30mK$. All data shown in this paper are qualitatively reproducible with thermal cycling.

\begin{figure}[htbp]
		\includegraphics[width=8.8cm, height=6.2cm]{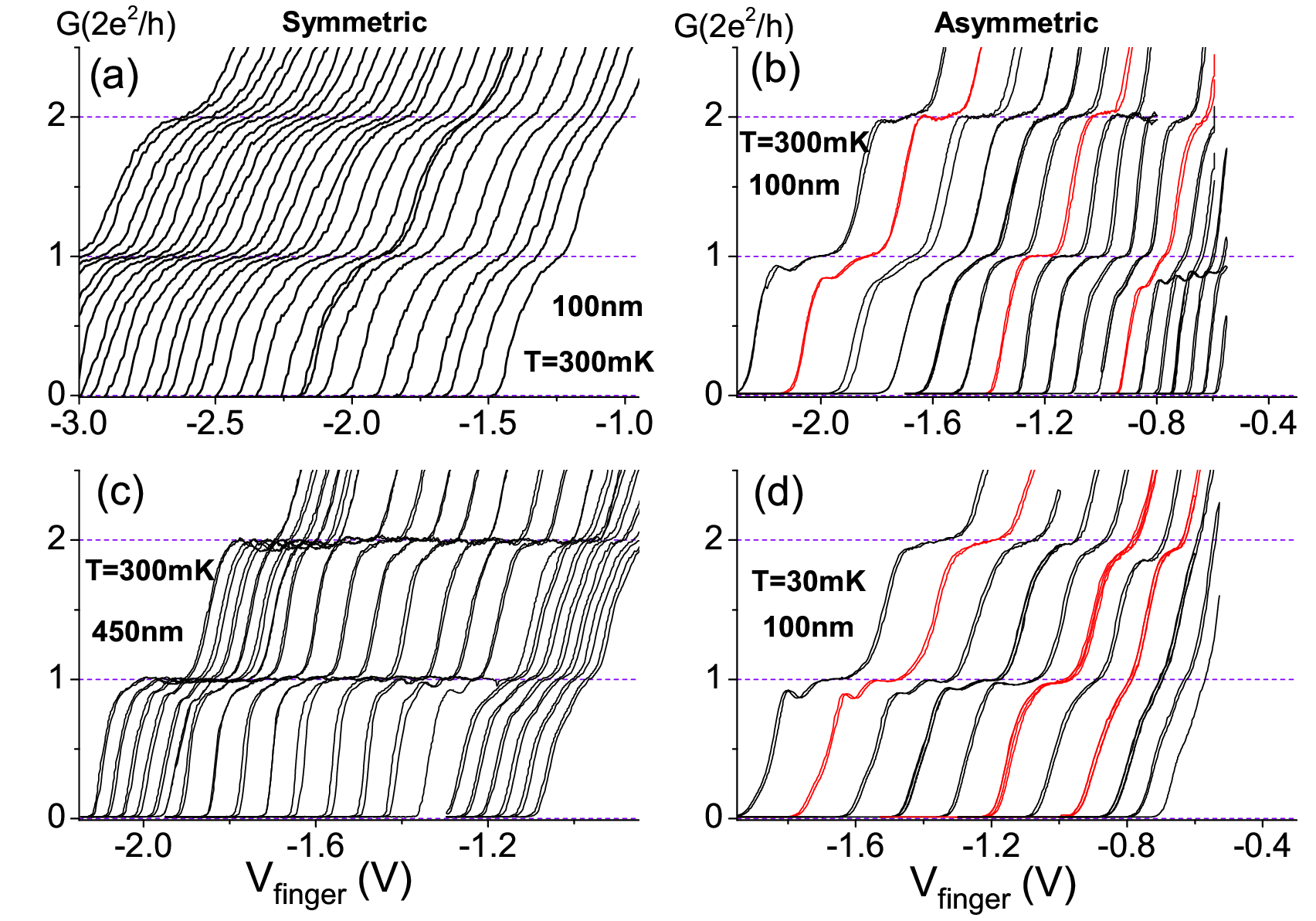}
		\caption{(Color online) Linear conductance (G) traces of QPCs with different gate geometry. (a),(c): symmetric QPCs with lithographic channel length 100 nm and 500 nm, with channel width, respectively, of 150 nm and 450 nm (450 nm at center, opening up parabolically to 680 nm at the ends), $T=300~mK$. (b),(d): the same asymmetric QPC with lithographic channel length of 100 nm and width of 250 nm at $300~mK$ (b) and $30~mK$ (d). Each trace is obtained by fixing a (wall) gate voltage ($V_{wall}$) while sweeping the other (finger) gate voltage ($V_{finger}$). $V_{wall}$ settings for the leftmost (rightmost) trace in (a)-(d) are, respectively, -2 V (-3.5 V), -0.8 V (-1.65 V), -1.84 V (-2.55 V) and -0.8 V (-1.3 V).}
		\label{linear}
\end{figure}

Before presenting the main data for the $dI/dV$ oscillations in the 100 nm channel length, asymmetric QPC, we begin by addressing the issue of donor/impurity-induced disorder in our QPC devices.  To establish with confidence that what we observe are intrinsic effects, and not from disorder, we contrast the behavior of the linear conductance in an asymmetric QPC, against symmetric QPCs with smooth and gradual entrances and exits. Fig. 1 shows the linear conductance for both symmetric (left panels) and asymmetric QPCs (right panels). Symmetric QPCs have two geometrically symmetric split gates while the asymmetric ones were fabricated by replacing one split gate with a long gate. This long gate is labeled the wall gate while the other short gate labeled the finger gate. The scanning electron microscope images of asymmetric QPCs can be found in Ref. \cite{Wu, Hao}. The left two figures in Fig. 1 show the linear conductance for two symmetric QPCs with different channel lengths (100 nm and 500 nm respectively), while the right two figures for one asymmetric QPC at $300~mK$ and at $30~mK$, respectively. The data in Figs. 1(b) and (d) for the 100 nm asymmetric QPC are from different cool downs. During measurement, the wall gate voltage for the asymmetric QPC ($V_{wall}$) was held fixed, while the $V_{finger}$ gate voltage was scanned to measure the conductance. 

In our previous work \cite{Hao}, conductance resonances and their behavior versus channel length were reported. The dirt effects, caused by dopant impurity and lithographic imperfections, were largely ruled out. Here using the symmetric QPCs as a control group, we present more clear-cut evidence to rule out dirt effects. (The supplementary document \cite{supplement} contains an expanded discussion.) Especially for Fig. 1(c), in a symmetric QPC with a long, $500~nm$ channel length, almost no conductance resonances are observed down to 300mK (except for the 0.7 effect \cite{Thomas}). While for the $500 nm$ long asymmetric QPC fabricated on the same crystal, dense conductance resonances are observed (in Fig. 2 (c) of Ref. \cite{Hao}). If the resonances in asymmetric QPCs were caused by dirty effects, then the symmetric QPCs (Fig. 1(a)(c)) should also show dirt-induced resonances. The absence of resonances in the symmetric QPCs other than the quantized plateaus clearly rules out impurity effects. Thus, the resonances are indeed intrinsic, which as ascribed, are due to electron correlations. Besides resonances, modulation of conductance plateaus was also observed in asymmetric QPCs.\cite{Wu, Hao} The symmetric QPCs with smoothed entrance/exit (Fig. 1(a)(c)) show no plateau modulation as the $V_{wall}$ was tuned from the left most to the right most, e.g. the value of plateau remains at the quantized value throughout, while for the asymmetric QPC (Fig. 1(b)(d)), clear modulation of the first and second quantized plateaus are observed. The dotted lines mark the quantized plateaus positions. 

\begin{figure}[htbp]
		\includegraphics[width=8.8cm, height=6.2cm]{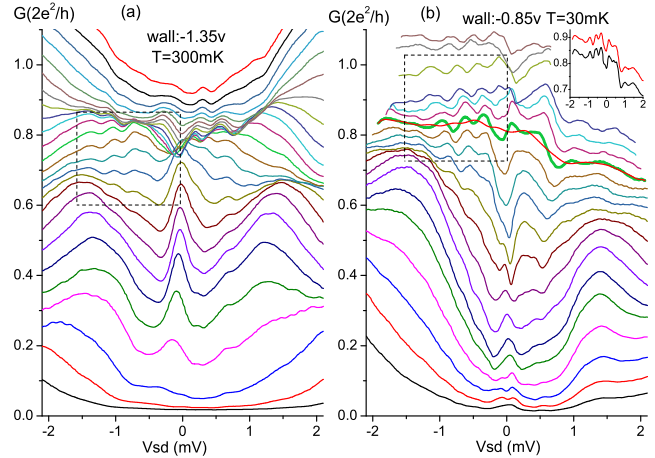}
		\caption{(Color online) $dI/dV$ as a function of source drain bias ($V_{sd}$) for fixed $V_{wall}$ and different $V_{finger}$ values (represented by different colors) (a): $V_{wall}$:-1.35v, $V_{finger}$: -0.89 V (bottom) to -0.71 V (top), $T=300~mK$. (b): $V_{wall}$:-0.85 V, $V_{finger}$: -1.84 V (bottom) to -1.65 V (top), $T=30~mK$. Inset: a typical $dI/dV$ curve for left and right sweeping $V_{sd}$, vertical offset by $0.06\cdot 2e^2/h$ for clarity. }
		\label{DIDV}
\end{figure}

Fig. 2 shows the $dI/dV$ as a function of source drain bias ($V_{sd}$), for the asymmetric QPC (corresponds with Fig. 1(b)(d)), at $T=300~mK$ (a) and $30~mK$ (b) for different cool downs. Unusual small $dI/dV$ oscillations, which have not previously been reported in QPCs, are developed. Several typical oscillations are highlighted in the regions enclosed by the black dashed rectangular boxes. We will focus on the systematic study of these $dI/dV$ oscillations in the remainder of this manuscript.

The $dI/dV$ measurement was configured in the following way: during the measurement, $V_{wall}$ was held fixed, while $V_{finger}$ was varied for different individual curves (different colors online). $V_{wall}$ in Fig. 2(a) was set at the same value as the rightmost red curve in Fig. 1(b), while Fig. 2(b) has the same $V_{wall}$ value as the leftmost red curve in Fig. 1(d).  The $dI/dV$ curves, corresponding with other red linear curves of Fig. 1(b)(d), were also systematically measured. These additional $dI/dV$ curves show very similar behavior with those presented in this paper (including their magnetic field and temperature dependence). The three linear red curves in Fig. 1(b) or (d) represent three linear conductance regions, which alternately exhibit a weak, strong and then weak first quantized plateau. The consistency of the three $dI/dV$ sets corresponding with the red linear curves in Fig. 1(b) or (d) for different cool downs, suggests that the oscillations exist in a wide gate voltage range. For each $dI/dV$ trace, $V_{sd}$ was swept back and forth at different speeds between $-2mV$ and $2mV$. The inset in Fig. 2 contains the back and forth sweeping for one typical curve to demonstrate reproducibility. 

\begin{figure}[htbp]
		\includegraphics[width=8.8cm, height=6.2cm]{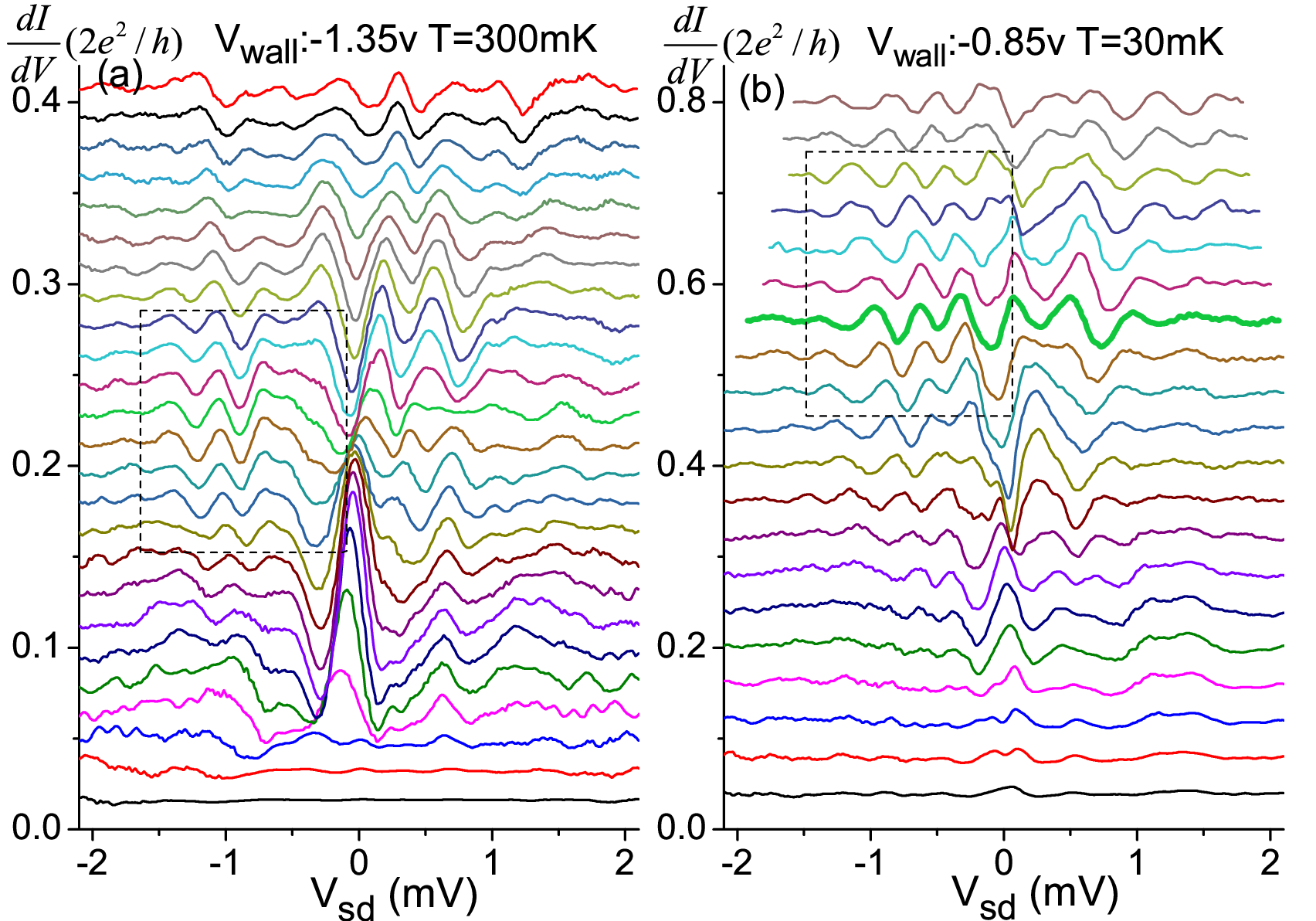}
		\caption{(Color online) $dI/dV$ curves of Fig. 2, after subtracting smoothed background. Vertical offset: 0.016 (a) and 0.04 (b) for clarity. The colored curves correspond to the curves in Fig. 2 of the same color. The curves enclosed by the black dashed rectangles in (a) and (b) correspond to the respective curves enclosed by the dashed rectangular boxes in Fig. 2(a) and (b), respectively. }
		\label{DIDVsub}
\end{figure}

To highlight these oscillations, a smoothed background was subtracted.  The background was generated by smoothing over a relatively wide $V_{sd}$ range to average out the small oscillations (the thin red curve, passing through the thick green curve in Fig. 2(b), represents a typical smoothed background of the thick green curve).  Fig. 3 shows the $dI/dV$ curves after background subtration, offset for clarity.  (Note that the thick green curve in Fig. 3(b) corresponds with the thick green curve in Fig. 2(b) after subtracting the red curve). A typical $dI/dV$ curve has around $5\sim 7$ oscillations. A typical oscillation amplitude at $T=30~mK$ is roughly $0.035\times 2e^2/h$. At $T=300~mK$, the oscillation size is roughly halved, as indicated by a comparison of the oscillations in the dashed rectangular boxes in Figs. 3 (a) and (b), where the vertical scale for (b) is twice as for (a). The separation between neighboring oscillations is $\sim 330\mu V$ in $V_{sd}$ at $T = 30~mK$ and at $300~mK$. 

\begin{figure}[htbp]
		\includegraphics[width=8.8cm, height=6.2cm]{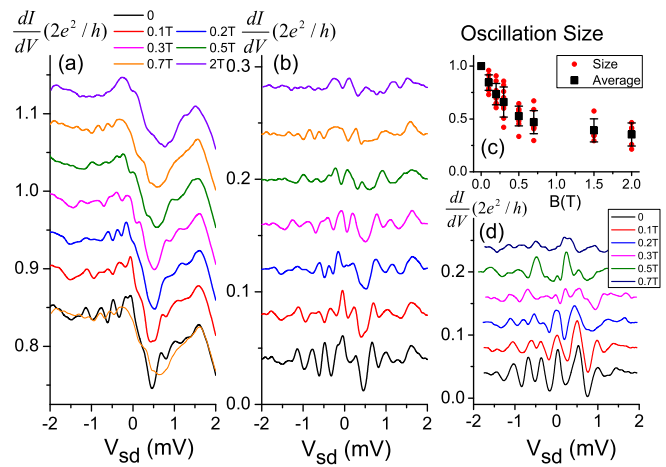}
		\caption{(Color online) The in-plane magnetic field dependence at $T=30~mK$ (a): $V_{wall}$: -1.1 V, $V_{finger}$: -1.08 V,  vertical offset: 0.05. The thin orange curve, going through the black curve at $B = 0~T$, is identical with the upper orange one ($B = 0.7~T$), but without the offset. (b) $dI/dV$ from (a) after subtracting the smoothed background, with offset: 0.04. (c) Statistics of the relative average oscillation amplitude for different gate voltages at different in-plane magnetic fields. (d) The $dI/dV$ with smooth background subtracted for a second set of voltages: $V_{wall}$: -0.85 V, $V_{finger}$: -1.725 V, offset: 0.04.  In (b) and (d), the positions of the oscillation crests hardly shift, when the in-plane field is increased to $0.1~T$ from $0$, with a concomitant out-of-pane component $\sim 0.0125~T$.  This observation is a strong piece of evidence, which helps rule out external interference paths as the cause of the oscillations, as 
detailed in the supplimentary document \cite{supplement}.}
		\label{B}
\end{figure}

Fig. 4(a) shows the in-plane magnetic field dependence of a typical $dI/dV$ trace. Different magnetic fields are represented with different colors. The curves in Fig. 4(b) correspond with those in (a) with the smoothed background subtracted. Fig. 4(d) shows the in-plane magnetic field dependence for a second set of fixed $V_{wall}$ and $V_{finger}$ values, (also with smoothed background subtracted). Figs. 4(b)(d) indicate that at a low in-plane magnetic field ($\sim 0.3~T$), the oscillation sizes are already significantly reduced, and are nearly fully suppressed by $\sim 0.7~T$. This surprising in-plane magnetic field behavior is systematically characterized and examined below.  
  
The magnetic field dependence of nine $dI/dV$ traces were measured to obtain reasonable statistics. These nine sets were chosen in the following way: for each red curve in Fig. 1(d), the magnetic field dependence of three $dI/dV$ traces, corresponding with three different $V_{finger}$ settings, were measured. Fig. 4(b) and (d) represent two of the nine $dI/dV$ sets. Fig. 4(c) shows a summary for the nine sets, providing a general trend of how the average oscillation size evolves with in-plane field. The average oscillation size was calculated in the following way: for each curve at a specific magnetic field, after subtracting the smoothed background, the average oscillation size was estimated by calculating the average absolute value for all data points of this new curve. Then we normalize the size against the size of each curve at $B=0~T$. For each field, there are nine red dots (some not visible) representing the average oscillation amplitude of each of the nine $dI/dV$ curves, while the black dot represents the mean value of the nine red dots. The fact that the oscillation amplitude does not decay fully to zero may be due to non-zero contributions not related to the oscillations, such as noise.  Another method, based on the average root-mean-square, instead of the average absolute value, showed similar trends as  Fig. 4(c). At $B=0.7~T$, the mean oscillation size (black dot) of these nine $dI/dV$ curves is nearly saturated in its decay trend. 

In Fig. 4(a), the orange curve overlapping the black curve (B=0 T), is identical with the upper orange curve (B = 0.7 T), but without offset. The orange curve, for which the oscillations are nearly suppressed, passes through the black curve for the most part. This behavior is found for all the other $dI/dV$ traces: these curves with reduced oscillations pass through the B=0 T curves, when plotted without vertical offset. This behavior will be discussed in the supplemental document \cite{supplement} as one evidence to rule out the weak/anti-weak localization scenario.  If we were to assume a typical g-factor of $-0.44$, the Zeeman energy scale would be $ 51 \mu eV/Tesla$, a factor of three smaller than than the oscillation separation($\sim 330\mu eV$) even at $B= 2~T$.  We emphasize that the sensitivity to such a small in-plane B field ($\sim 0.3 T$) is highly unusual and completely unexpected!

\begin{figure}[htbp]
		\includegraphics[width=8.8cm, height=6.2cm]{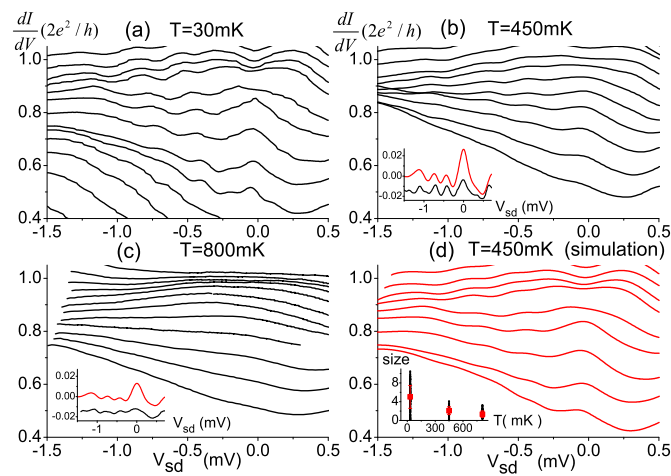}
		\caption{(Color online) Temperature evolution of the small $dI/dV$ oscillations. $T=30~mK$, $450~mK$, $800~mK$ for (a)(b)(c) respectively. (d) is the numerical simulation for T=450 mK. Inset in (b)(c): one typical $dI/dV$ (black) and numerical(red) curve after subtracting smoothed background, for $T=450~mK$ (b) and $800~mK$ (d), respectively. Inset in (d): statistics for the average oscillation amplitude for different gate voltages at different temperatures.}
		\label{T}
\end{figure}

Fig. 5 shows the temperature evolution of these $dI/dV$ oscillations, for which the $V_{wall}$ corresponds to the middle red linear conductance curve in Fig. 1(d), at $T=30~mK$ (a), $450~mK$ (b) and $800~mK$ (c). The oscillations become smaller at higher temperatures and are almost washed out at $800~mK$. The temperature evolution of the oscillations, corresponding to the right and left red curves in Fig. 1(d), show consistency with Fig. 5. The $V_{wall}$ in Fig. 5(a) is held fixed at $-1.05~V$, while at higher temperatures($450~mK$, $800~mK$), due to drift, and by comparing the linear conductance traces, $V_{wall}$ was fixed at $-1.10~V$ in Figs. 5(b) and (c).  

Fig. 5(d) shows the numerical simulation of the $dI/dV$ at $T=450~mK$, which qualitatively agrees with the measured data in Fig. 5(b). The simulation, based on thermal broadening of the Fermi distribution, involves convoluting the transmission probability with the Fermi function to deduce the conductance at higher temperatures, as discussed in the supplemental document \cite{supplement}.  

To better compare the measured oscillations with the simulation at each temperature, in the insets in Fig. 5(b)(c), we show the $dI/dV$ after subtracting the smoothed background, for one representative trace in Fig. 5(b) or (c). By comparing the measured data (black) and the simulated ones (red) at $T=450~mK$ and $800~mK$ in the inset of (b)(c), the oscillation sizes are seen to be semi-quantitatively in agreement, for the oscillations off zero bias. However, the structure at zero bias does not agree with, and is obviously smaller than, the simulation results. This difference in behavior suggests that the two bias regions may have different energy scales and thus are related to different origins, e.g. multi-Kondo scales or a new mechanism.  In such a Kondo scenario, it is still single channel physics, as opposed to 2- or higher channel.  Thus, the behavior should still be Fermi-liquid-like.

The inset in Fig. 5(d) shows the statistics for the average oscillation size of every $dI/dV$ curves at different temperatures. This statistics provides a trend of how the oscillation sizes decreases as increasing the temperatures. The oscillations are almost washed out at $T = 800~mK$. The corresponding thermal energy scale, is $2.6~\mu eV$, $39~\mu eV$, and $69~\mu eV$, respectively.  The average oscillation size is calculated using the same method discussed in the statistics of magnetic field dependence. At a given temperature, the black vertical line indicates the spread in the average oscillation size for individual curves at a particular gate voltage, while each red dot is the mean value of all curves contained within the black line.

In the above, we have provided a fairly complete characterization of the oscillations.  At present, there does not exist any theory, which can be used to interpret the unusual $dI/dV$ oscillations, although a high-spin Kondo effect reflecting an underlying electron lattice represents an important possibility. Several possibilities that may arise within simple single particle scenario are considered in the supplemental document and ruled out \cite{supplement}, due to the fact that the size of the oscillations and its magnetic field dependence behavior do not agree with any of these models: for example, the contribution from the 2DEG reservoir can only give rise to an oscillation with a size much smaller than measured. Thus the oscillations are from the QPC channel. The possible excitation to quasi-bound states, modulation of energy bands due to magnetic field, field misalignment, Aharonov-Bohm effects and weak or anti-weak localizations, are also carefully considered, but are inconsistent with the data (and its magnetic field dependence). 

By virtue of having ruled out all sensible single particle pictures, we are led to the conclusion that our observation must be related to electron-electron interactions. Since an in-plane magnetic field mainly interacts with electron spin, the magnetic field dependence suggests that the small oscillations are related to electron spin correlation. Based on the estimated 1D electron density in the single channel limit \cite{Wu_thesis}, a rough estimate of the electron number inside the QPC channel is around $3\sim 6$, which is of the same order as the number of oscillations for a typical $dI/dV$ curve. In future studies, it will be interesting to directly relate such oscillations to the formation of localized charge state inside the QPC channel due to correlation effects, possible of relevance to the exotic zigzag type of Wigner-crystal-like states \cite{Matveev, Abhijit}, for instance, via the manifestation of a high spin Kondo state, reflection an underlying lattice 
of spontaneously formed, localized electron array.

We thank M. Melloch for the GaAs/AlGaAs crystal. This work was supported in part by NSF DMR-0701948, and by the Academia Sinica, Taipei.

\end{document}


\title{Novel Differential Conductance Oscillations in Asymmetric Quantum Point Contacts (Supplemental)}
\author{Hao Zhang}
\affiliation{Department of Physics, Duke University, Physics Building, Science Drive, Durham, North Carolina 27708, USA}
\author{Phillip M. Wu}
\affiliation{Department of Applied Physics and Geballe Laboratory for Advanced Materials, Stanford University, Stanford, California 94305}
\author{A. M. Chang}
\affiliation{Department of Physics, Duke University, Physics Building, Science Drive, Durham, North Carolina 27708, USA}

\date{\today}

\maketitle

\section{Argument For Intrinsic Behavior}

We address the key issue as to whether the characteristics we have uncovered are intrinsic to the QPC geometry and to interaction-induced correlations, or are instead, the much less interesting consequences of undesired disorder.  In particular, we wish to highlight in detail, the entire set of evidence indicating that these represent fundamentally new behaviors.  

Short of directly imaging the potential landscape--a capability unavailable at present, there is no direct method to prove that the observed features are intrinsic, i.e. due to geometry and interaction only .  However, the characteristics we have observed, taken in totality, point clearly toward intrinsic behavior  rather than disorder-induced behavior.  Many of these characteristics are unique, in the sense that we are unaware of other systems, disordered or otherwise, which exhibit such characteristics.

First, regarding the occurrence of the sharp resonance themselves, the statistics of their observability points toward intrinsic behavior.  
Here we summarize our finding: (1) Out of more than 20 asymmetric QPCs studied, all exhibit a modulation of the first quantized conductance plateau, as the wall gate voltage is systematically stepped, as can be seen in Fig. S-1 (c) - (h).  For certain gate settings, a suppression of the $1*\frac {2e^2}{h}$ plateau can be found in each panel. (2)  Of these 20 asymmetric devices, all but 1 exhibit sharp resonances at $300~mK$ temperature.  (3)  Of all 8 symmetric QPCs studies, all show either no resonances, or considerably fewer and/or broader ones, with only 1 exception, for which unevenness in the lithography is suspected to cause excessive scattering.  (4)  In the smooth and short symmetric QPCs, for all 3 $100~nm$ channel length QPCs, there are NO visible resonances
or modulation in the first plateaus at all.  In Fig. S-1(a), we present a typical example.  Moreover, a $500~nm$ channel length,
smooth QPC (geometry: $450~nm$ gap at the middle, and opens up parabolically to $680~nm$ gaps at the two ends), exhibits the same behavior, as shown in Fig. S-1(b).  If disorder or lithographic imperfections were dominant, these devices should exhibit sharp resonances as well.  (5)  In complete contrast, Fig. S-1(c)-(h) indicate that for asymmetric devices, with channel length ranging from $100~nm$ to $700~nm$, sharp resonances and modulation of the first plateau are ubiquitous. The progression in the plateau modulation is very orderly,  with little hint of randomness.  The general features are reproducible upon thermal cycling to room temperature!

\begin{figure}[htbp]
		\includegraphics[width=8.8cm]{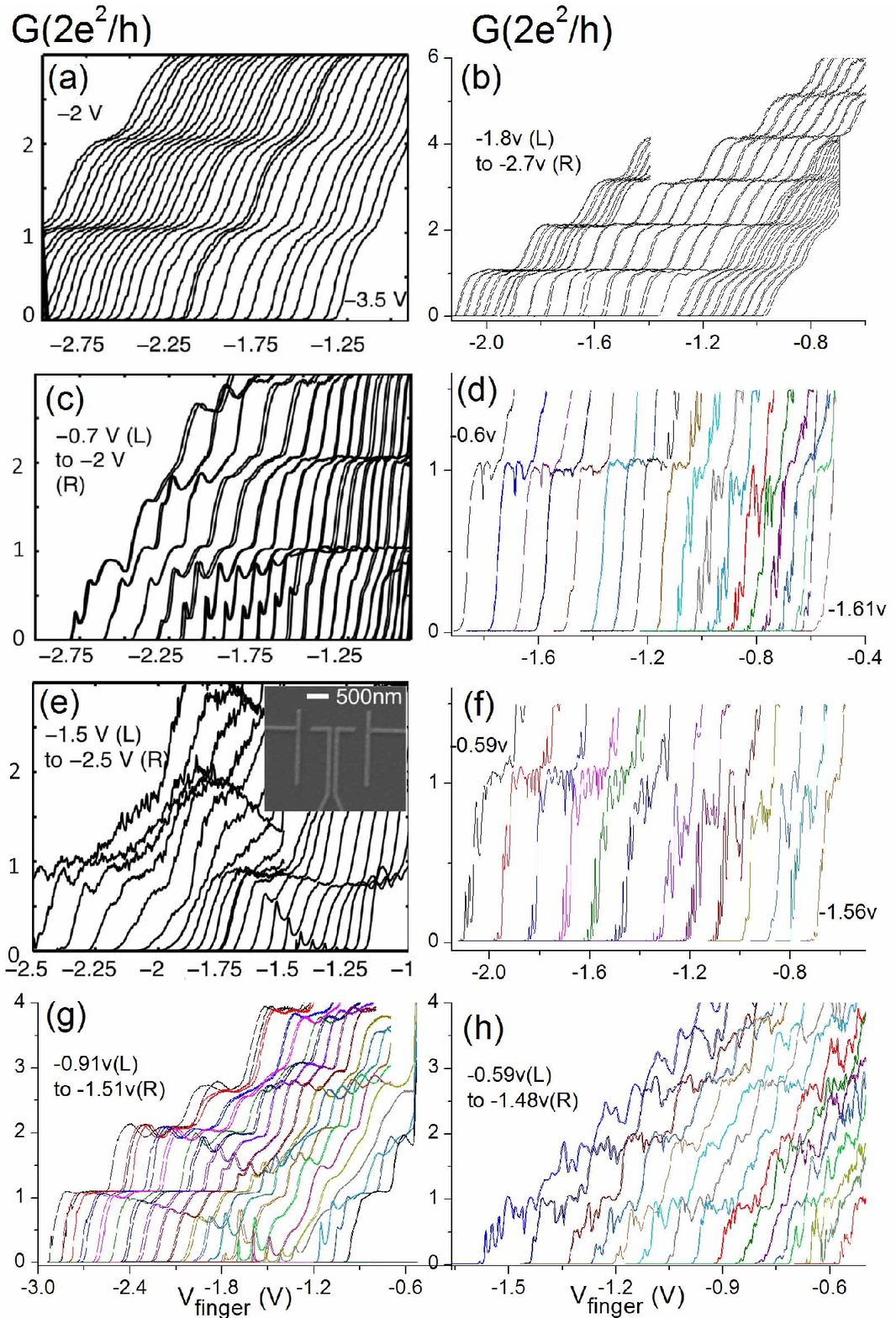}
		\caption{(Color online) (a)-(b) conductance of smooth, symmetric QPCs with $100~nm$ channel length (a), and $500~nm$ channel length (b).  For the latter, the exact geometry is described in the text.  (c) - (h) asymmetric QPCs of short, $100~nm$ channel length [left panels
(c), (e), (g)], and long $500~nm$ channel length [right panels (d) and (f)] and long $700~nm$ channel length [right panel (h)].
The temperature is $300~mK$ for all traces.  Note that the behavior of the symmetric channels [(a) and (b)] are dramatically different from the asymmetric ones [(c) - (h)]. In particularly, they do not exhibit any resonance or modulation of the first conductance plateau, in complete
contrast to ubiquitous presence of such features in the asymmetric QPCs.  The behaviors of the symmetric QPCs are indicative that disorder plays a minor role in producing the features we observe in our asymmetric devices.}
\end{figure}

Although we have only studied one $500~nm$ long channel symmetric QPC, for which no resonances or modulation of the conductance plateau position are observed, and thus the statistics is not compelling, we stress that this is highly unusual.  Even in symmetric QPSs longer than $100~nm$ in channel length, we invariably see resonances, although there are generally significantly fewer of them, and they tend to develop 
at lower temperatures.  Therefore, having no resonances/plateau modulation is extremely unusual.  For the device shown in Fig. S-1(b), 
only by deliberately smoothing the entrance and exit, were we successful in ridding of all resonances.  Other symmetric QPSc written on the same chip, but without the fully smoothed entrance/exit, all showed resonances/plateau modulation to varying degrees.  In short, if disorder 
caused the resonances, it is highly unlikely that any channel can be free of resonances, particular for one that is $500~nm$ in length.  
For reference, the lateral spacing between ionize silicon donors is approximately $15~nm$, considerably shorter than the shortest channel length of $\sim~100~nm$!


Second, the quasi-period small oscillations in the differential conductance $dI/dV$, as a function of source drain voltage $V_{sd}$, the focus in the main text of this paper, cannot be accounted for within a simple, single particle scenario (see the following sections in this supplemental document).  Moreover, we are unaware of any conventional system, ordered or disorder, in which interaction 
is known to induce such oscillations.


Third, in our $100~nm$ long asymmetric QPC devices, certain traces exhibit no odd $1 \times e^2/h$ plateau at all, even in the presence of an 
in-plane B-field up to $9.1 T$ down to $30~mK$ in temperature! In the absence of electron correlation, the g-factor for an electron 
is $- 0.44$, leading to a clearly observable set of odd plateaus at in-plane fields as low as $4~T$ \cite{Cronenwett}.  
The absence of the $1 \times e^2/h$ plateau can only arise from an usual electron correlation effect.

Fourth, the zero bias anomaly is also unusual at high in-plane field.  Above $B \sim 5 T$, a very distinct, small cuspy feature is observable at many voltage settings.

The findings that we have detailed here are difficult to explain in terms of disorder.  Further, they do not conform to the expectations of a disorder system in the presence of strong Coulomb interaction within any of the conventional scenarios, such as the Fermi liquid theory, simple Luttinger liquid theory, or Coulomb gap theory.  Therefore, we feel that there is strong argument that these findings are likely the 
consequence of novel correlations within the 1D QPC channel.

\section{Numerical Simulation for the Temperature Dependence of the Oscillations in Fig. 5 of the Main Text}

The simulation is done in the following way: since the short QPC channel( $\sim 100nm$) is coupled to the two Fermi sea reservoirs, the QPC transport will be dominated by the Fermi statistics of these two reservoirs. Thus the conductance will be the integral of the product of electron transmission probability ($T(E)$), derivative of Fermi-Dirac function ($\frac{\partial f}{\partial E}$), density of states and velocity as a function of energy, for both left and right Fermi sea reservoirs. The expression for the differential conductance is obtained and shown in the following equation:
\begin{equation}
	\resizebox{0.9\linewidth}{1.1\height} {$\frac{dI}{dV}=\frac{2e^2}{h}\int_{-\infty}^{\infty}\big[ \alpha \frac{\partial f(E-\mu_L,kT)}{\partial E} + \beta \frac{\partial f(E-\mu_R,kT)}{\partial E} \big] \scalebox{.7}{$T(E)dE$}  $}
\end{equation}
 where $\mu_L$ and $\mu_R$ are the quasi-Fermi level for the left and right reservoir, $\alpha$ and $\beta$ represent the portion of bias distributed between the left and right reservoir respectively. Thus $\alpha+\beta=1$. During the measurement, one reservoir is virtually grounded, leading to $\alpha=1, \beta=0$. The simulation shown in Fig.5 is based on this assumption. At $T=30mK$, the width of the derivative of Fermi-Dirac function, $3.5kT\approx 9\mu eV$, is much smaller than the width of the small oscillations ($\sim 330\mu V$), thus to the first order approximation, the $dI/dV$ curves at $T=30mK$ can be treated as if it was at $T=0K$. At $T=0K$, the derivative of Fermi-Dirac function in Equation 1 becomes a delta function, thus $T(E)$ can be extracted. Using the $dI/dV$ curves at $T=30mK$ as an input to extract $T(E)$, the high temperature evolution can be simulated by plugging in $T(E)$ in Equation 1. One typical simulation for $T=450mK$, using Fig.5(a) as an input, is shown in Fig.5(d). As can be seen, the oscillation sizes qualitatively agree with the measured $dI/dV$ at the same temperature (Fig. 5(b)). The simulation for $T=800mK$ (not shown here) also qualitatively agrees with Fig.5(c). Besides this set, the simulation for other sets of $dI/dV$ curves, with the wall voltage corresponding to the leftmost and rightmost red curves in Fig. 1(d), show consistency with the measured $dI/dV$ oscillations at $T=450mK, 800mK$. Besides the assumption $\alpha=1, \beta=0$, simulations based on $\alpha=0.9, \beta=0.1$ and  $\alpha=0.7, \beta=0.3$ are also implemented and show similar results compared with the $\alpha=1, \beta=0$ case.

\section{Discussion of Simple Single-particle Pictures}

We first rule out any artifacts from either the indium contacts, or from the two-dimensional electron gas (2DEG) 
regions leading into the QPC, as we discuss below.  Then, we will present evidence, based on the 
lack of fringe shift (shift in the position of oscillations) in a magnetic field producing an Aharonov-Bohm phase, 
to rule out a Fabry-Perot interferometric effect from the presence of an accidental scatterer outside the QPC.  
Consequently, we can ascribe the observed novel behaviors to the 1D QPC channel with full confidence.  

{\bf Indium Contacts, and the Two Dimensional Electron Gas (2DEG)}

Let us begin by examining the issue of contact resistance.  Typical good quality indium contacts to the 2DEG are below $1 \Omega$ at low temperature.  This magnitude is far smaller than the observed oscillation amplitude, which is as large as $200 \Omega$.  Therefore, artifacts from the indium contacts can be safely ruled out.  As far as the 2DEG regions, which behave as excellent metallic conductors, the only possibility for any variation in the resistance (or conductance) in this temperature range ($30 mK - 1 K$) comes from quantum interference effects, such as Aharonov-Bohm interference, or weak localization/anti-localization.  Given the long $l_{\phi}$ typically exceeding $10 \mu m$ at $30 mK$, we expect such effects to be of order $G_0 = 2e^2/h$, compared to the 2DEG conductance of $G_{2D} \approx 60 G_0$ (equivalent to $\sim 220 \Omega$).  A change of order $G_0$ in the 2DEG will yield a corresponding resistance change of $\sim 6 \%$, or $13 \Omega$, far below the observed resistance modulation of $200 \Omega$.  Therefore, the 2DEG regions cannot account for our observations.  These considerations give substantial confidence that our findings are associated with the 1D channel.

{\bf Single Particle Scenarios inside of the Quantum Point Contact}

Next, we examine several single particle effects occurring within the quasi-1D QPC channel, including: (1) excitation into higher longitudinal bound states associated with quantization along the length of the channel, (2) misalignment of the in-plane magnetic field ($\sim 7^o$), giving rise to a vertical component out of the 2DEG plane, and (3) quantum interference, weak-localization/antilocalization, and Aharonov-Bohm interference type effects.  We argue that none of these scenarios is consistent with observations.

One possible origin causing these oscillations may be the excitations to the longitudinal quasi-bound states formed in the channel, associated with the channel length. If this were the case, the effective wave number and channel length (100nm) for the Nth quasi-bound states should satisfy $k_N\cdot L=N\pi$, similar with the relation of a $Fabry-P\acute{e}rot$ interferometer. Thus even the smallest energy level spacing, which is between N=1 and N=2, is estimated to be $\Delta E=\Delta\frac{(\hbar k)^2}{2m^*}=1.7meV$. This energy level spacing is 5 times greater than the spacing between the neighbor oscillations, which is around $330\mu V$ in $V_{sd}$. Besides this level spacing, a small in-plane magnetic field $\sim 0.3T$ is expected to do little to the electron transport in this quasi-bound state excitation picture. But the $dI/dV$ oscillations were significantly modified (suppressed) under such small in-plane magnetic field. These considerations suggest that the oscillations cannot be due to this excitation into higher single particle quantum levels.

Now we estimate how the magnetic field modifies the potential profile of the QPC in the 2DEG. We examine 
the effect of sidewall confinement, in the presence of an out-of-plane magnetic field, which is presence 
due to the $7^o$ misalignment of the in-plane field.  Our consideration is, in essence, the quantum-mechanical version 
of the suppression of the cyclotron motion by the lateral confinement potential.  
For illustrative purposes, we assume the confinement profile to be of a parabola shape: $\frac{1}{2}m\omega ^2 x^2$ along the direction perpendicular to the channel (out-of-plane). The $\omega$ was estimated to be $3.14\ast 10^{12}/s$,\cite{Wu_thesis} using the magnetic depopulation method.\cite{Magnetic_depopulation} Applying a magnetic field perpendicular to the 2DEG will hybridize this parabola with Landau levels to a new parabola with a new $\omega$. Since the in-plane magnetic field was aligned with the 2DEG within $7^{\circ}$, the modification of $\omega$ due to this out-of-plane magnetic field hybridization at $B = 0.3 T$ is estimated to be less than $0.2\% {\bf \cdot} \omega$. Such a small modification of the potential profile has little effect on the electron transport, as the effective $\omega$ increases quadratically with B. Thus this single particle picture cannot give rise to the significant modulation of the small oscillations either. 

Another way to interpret these oscillations is to relate them with the Aharonov-Bohm effect within the QPC channel. But the estimated Aharonov-Bohm phase shift at $B=0.3T$, due to the out-of-plane magnetic field component, is $\Delta \varphi=\frac{e\Phi_B}{\hbar}\approx 0.05~radians$, which is much too small to affect the transport. In this estimation, to calculate $\Phi_B=B_{\perp} {\bf \cdot} A$, where A is the effective area of the QPC channel, the channel length is assumed to be the same with the lithographic length, 100nm, and the effective channel width is assumed to be $10\%$ of the estimated 80 nm channel width in the single channel limit \cite{Wu_thesis}; within a ballistic channel, the fact that the wave function phase does not sample across the entire physical channel width gives rise to this reduction factor, in contrast to a diffusive case.\cite{Chang_AB} 

In the above, we have already established that Fermi-Dirac smearing is able to account for the reduction of the small oscillations amplitude with increasing temperature.  We have also argued, based on the magnetic field dependence (and accounting for a small 7$^o$  misalignment with
the 2DEG plane), that quantum interference/weak localization type mechanism is unable to account for them.  Here, we further point out that in the unlikely event that quantum interference or weak localization becomes of  relevance, any issue of the quantum phase coherence length,
$l_{\phi}$, should not lead to discernible effects, while  further emphasizing the unlikelihood of the quantum interference scenario.  

At low temperatures, at low energies $l_{\phi}$ typically exceeds $20 \mu m$ below $50 mK$ for high mobility ($\sim 10^6 cm^2/Vs$) devices \cite{Chang_WL}.  The temperature evolution of $l_{\phi}$ is typically power law, behaving as $T^{-\frac{1}{3}}$ or $T^{- \frac{1}{2}}$, depending on whether it is limited by small energy transfer or large energy transfer.  Even at $450 mK$, $l_{\phi}$ is expected to exceed $5 \mu m$, still far longer than the QPC channel length of $\sim 100 nm$.  Therefore, there should be little influence on transfer in the linear regime (low energies) due to weak/anti-weak localization. Moreover, the $dI/dV$ oscillations occur at a source drain bias voltage of $\sim 330 \mu eV$ and higher, far exceeding the thermal energy at $800 mK$ and below. Therefore, the relevant $l_{\phi}$ at such high energies should be completely insensitive to temperature in the range of our study. Besides that, for weak/anti-weak localization, the conductance at a finite magnetic field is either higher or lower than the zero magnetic field case, but as shown in Fig. 4(a) of the main text (the orange curve online), the conductance at a finite magnetic field goes through the zero magnetic field case, suggesting that it is not weak/anti-weak localization.

\begin{figure}[htbp]
		\includegraphics[width=8.8cm]{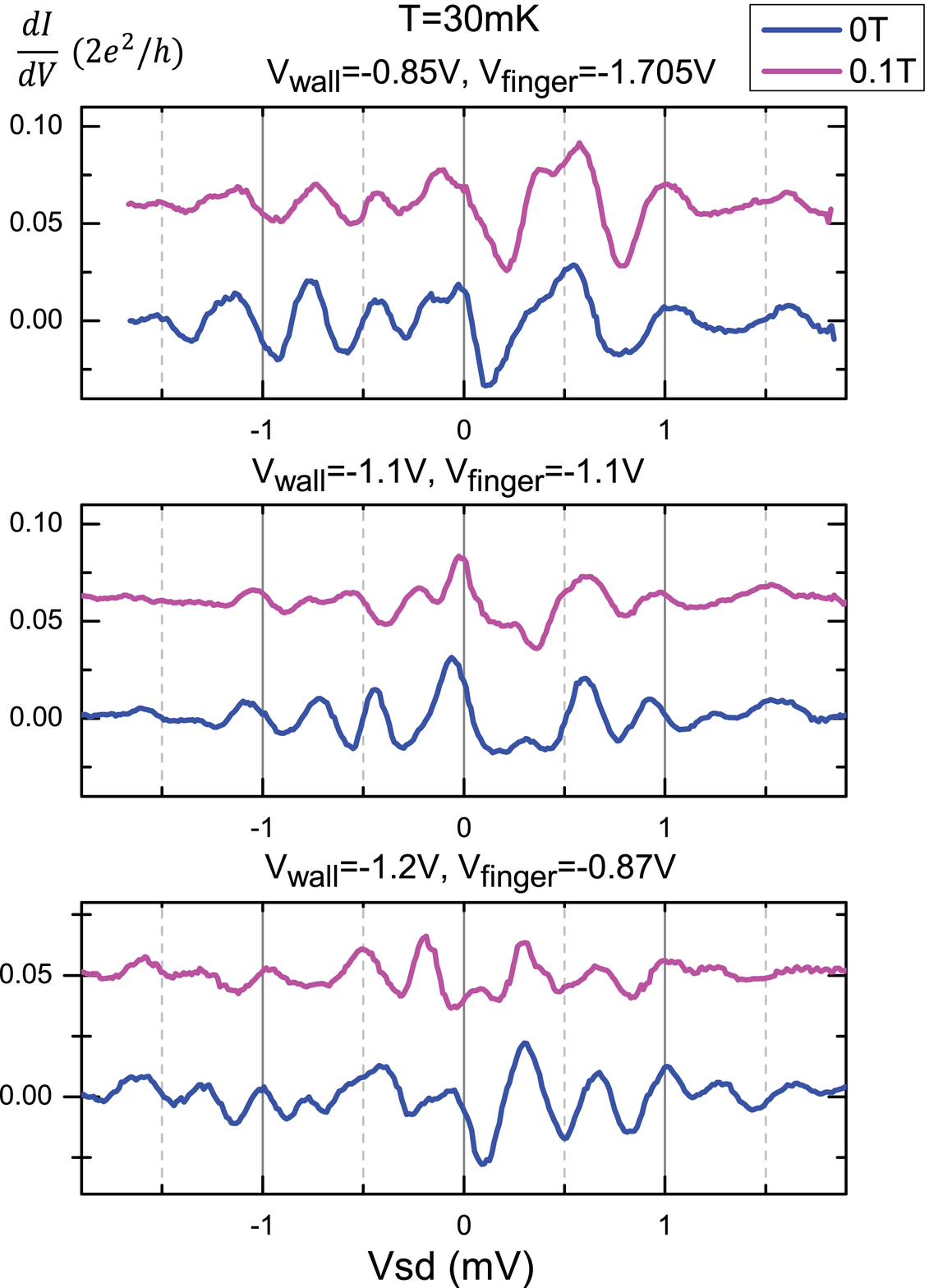}
		\caption{(Color online) dI/dV of the 100nm channel asymmetric QPC as a function of $V_{sd}$, at different $V_{wall}$ and $V_{finger}$, and B=0T and 0.1T in-plane magnetic field.}
\end{figure}

\section{Fabry-Perot Interferometer due to an Accidental External Scatterer and an Estimate of the 
Enclosed Aharonov-Bohm Flux for $B_{\perp} = 0.0125 T$ }

Last, and perhaps most importantly, we rule out the possibility of an accidental localized defect/scatterer within the 2DEG, in close proximity to the QPC, which, together with the entrance (or exit) of the QPC, act as a Fabry-Perot type interferometer.  The key to ruling out such a mechanism, which is based on ordinary quantum interference, is the {\it absence of a systematic shift} in the fringes 
(position of the $dI/dV$ oscillations versus source drain bias $V_{sd}$), in the presence of an Aharonov-Bohm flux.  This is best illustrated by comparing the data at $B = 0 T$ to those at $B = 0.1 T$.  Due to the $\sim 7^o$ misalignment of this in-plane field, an out-of-plane, 
perpendicular component $B_{\perp}$ is present in the latter case.  The data for three wall settings ($V_{wall}$) are presented in Fig. S-2.  The data indicate that out of approximately 25 oscillation crests, roughly 75 \% of them shift less than $\frac {\pi}{4}$.  Note that we 
present 1/3 of the total data.  The additional data are for 2 additional, slightly different values of $V_{finger}$ at each $V_{wall}$ setting.  The difference is relatively minor from the traces shown.  The clear-cut absence of shift for a vast majority of the oscillations unequivocally rules out such a scenario, in which an accidental scatter outside of the QPC produces a Fabry-Perot type interferometer.

Below, we provide a quantitative estimate of the Aharonov-Bohm phase shift based on the trajectory shown in Fig. S-3.  There, we find 
that the expect shift is at least $\pi/2$, assuming that the Aharonov-Bohm phase shift is the only contribution to the phase change.  If we included the increase path length due to the bending of the paths from the cyclotron motion, this estimated shift increases to $\sim 0.88 \pi$.  


The device parameters needed are the carrier density in the 2DEG ($3.8 \times 10^{11} cm^{-2}$), the corresponding Fermi ebergt ($E_F \approx 13.1 meV$), Fermi wave number ($k_F \approx 1.545 \times 10^6 cm^{-1}$), Fermi velocity ($v_F \approx 2.67 \times 10^7 cm/s$), effective
mass ($m* = 0.067 m_e$, where $m_e$ is the electron mass), and oscillation period in energy units ($\Delta E = e \Delta V_{sd} \approx 330 \mu eV$).

Based on the above, the change in $k_F$ per oscillation is given by:
\begin{equation}
\Delta k_F = \frac {d k_F}{d E} \Delta E
        = \frac {2}{\hbar v_F} e \Delta V_{sd}
        \approx 3.75 \times 10^4 cm^{-1}.
\end{equation}
Each oscillation involves a change in phase of $2\pi$. Thus, the path length L is given by:
\begin{equation}
\Delta k_F L = 2 \pi ,
\end{equation}
or $L \approx 1.67 \mu m$.

In the presence of an out-of-plane magnetic field of $B_{\perp} = 0.0125 T = 125 Gauss$, the cyclotron
radius is:
\begin{equation}
R_c = \frac {mvc}{eB} \approx 8.13 \mu m.
\end{equation}
the curved path (depicted by one of the two black arcs
in Fig. S-3) subtends an angle $\theta$:
\begin{equation}
\theta = 2 \sin ^{-1} (l/2R_c) \approx 0.2058 rad.
\end{equation}
The enclosed area is thus:
\begin{equation}
A = 2 * R_c^2 (\theta/2 - \sin \frac {\theta}{2}
        \cos \frac {\theta}{2}) \approx 0.0958 \mu m^2.
\end{equation}
One Aharovon-Bohm flux $\phi_o = \frac {hc}{e} = 41.3 Gauss-\mu m^2$. Thus, at $B_{\perp} = 125 Guass$, we have a flux $\Phi$:
\begin{equation}
\Phi = B_{\perp} A \approx 12.05 Gauss-\mu m^2 = 0.29 \phi_o ,
\end{equation}
or just over a $\frac {1}{4}$ cycle (just over $\frac {\pi}{2}$ in radians).

\begin{figure}[htbp]
		\includegraphics[width=8.8cm]{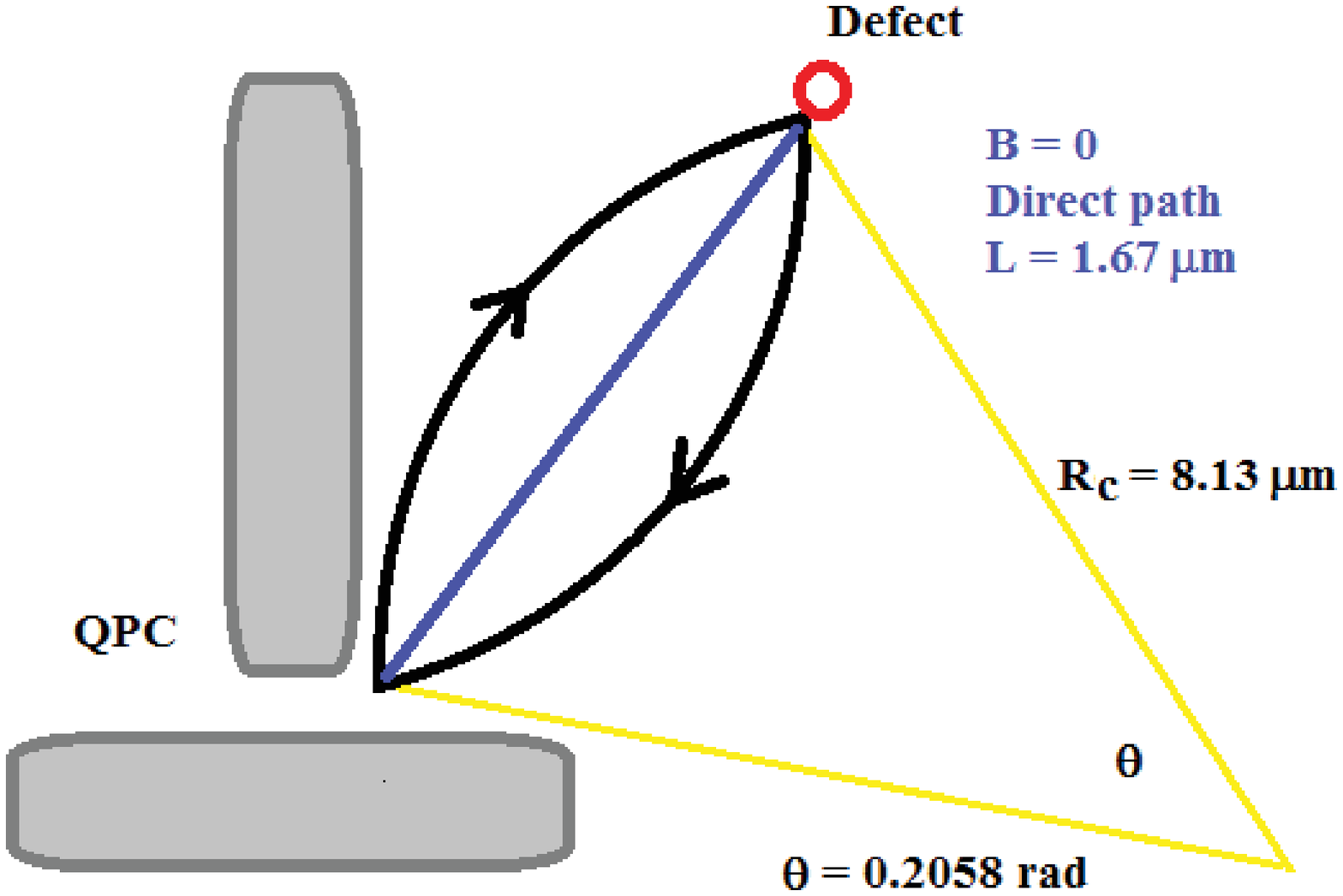}
		\caption{(Color online) Assume typical geometry and interference path between 
the entrance/exit of the QPC and an accidental, strong point scatter.  The curved paths indicate 
the cyclotron paths in the presence of a weak out-of-plane magnetic field.}
\end{figure}

In estimating the AB shift an assumption was made to neglected another contribution from the path length increase from the cyclotron bending.  At B = 0.1 T ($B_{\perp} = 0.0125 T$), in addition to an enclosed flux, the path lengths also increase slightly.  This changes the phase
by $\sim 0.29 \pi$.  If one were to add it to the $0.58 \pi$ from the AB, the shift would be $0.88 \pi$.

However, if they act in opposite directions, then it would only be $-0.29 \pi$.  In Fig. S-2, roughly 40 percent of the oscillation crests clearly shift less than this. 

There is a way to tell which way it should shift.  All the slight shifts in the data (excluding the few ones that shift a lot) are toward the right (i.e. toward more positive $V_{sd}$).  On the other hand, since electrons have negative charge, the higher chemical potential occurs on the negative $V_{sd}$ side, with a corresponding increase in reservoir $E_F$ by $-eV_{sd} > 0$.

This fact tells us that the fringes would shift toward the right, from the slight increase in path length arising from cyclotron bending; in other words, for a fixed phase $\phi = k_F L$, an increase in $L$ is compensated by a corresponding decrease in $k_F$, causing a shift toward lower $k_F$, and thus lower chemical potential in the 2DEG (i.e. to the right in $V_{sd}$).  

Since all the slight shifts in our data are to the right, if the Aharonov-Bohm and path length shifts acted oppositely, Aharonov-Bohm would win as it has the larger magnitude ($0.58 \pi$ versus $0.29 \pi$), and the resultant combined shift should be toward the LEFT, not toward the right.  Therefore, one is forced to conclude that they must be acting together, in the same direction toward the right side of the $V_{sd}$ axis, and the expected shift is the sum of the two contributions, equalling $0.88 \pi$, even larger than the initial estimate of $\sim 0.58 \pi$.  This expected shift (either $0.58 \pi$ or $0.88 \pi$) based on the interference scenario is at variance with most of the oscillation crests ($> 75 \%$ of the crests), and enables to rule out this mechanism of an interference path due to an accidental external scattering with substantial confidence.

Despite the above analysis, which attempted to include the contribution from the path length increase, in the end, the initial estimate, without this added contribution from the path length increase, may actually be correct.  This is because Onsager symmetry guarantees 
that for a 2-terminal measurement, $G(+B)~=~G(-B)$ at zero bias, with $G \equiv dI/dV|_{V_{sd}=0}$.  This means that even if we kept 
$k_F$ constant, but make a longer path, at $B = 0$, we always have either a maximum, or a minumum.  If we could assume that an extremum in B implies an extremum in $k_F$ (the latter is varied via $V_{sd}$), then at $B=0$, the phase is pinned despite an increase in path length. Since the path length change at 0.1 T only changes the phase by less than $2\pi$, it is reasonable to assume that the extremum does not flip from a maximum to a minimum.  This consideration means that even though there is a path length change, this change should NOT effect the phase of the fringes, and only the Aharonov-Bohm phase matters.  As a last and essential point, it is important to point out that in all previous studies of Aharonov-Bohm/quantum interference oscillations in a ring geometry, all the oscillations shift position in the presence of an Aharonov-Bohm 
phase, with no known exceptions.

\section{Magnetic Field Dependence of the Oscillations at 400 mK }

Fig. S-4 shows an example of magnetic field dependence of the oscillations, when the temperature was increased from 30mK to 400mK. As can be seen, at higher temperature the oscillation amplitudes become smaller, but the magnetic field dependence shows similar trend compare with the data in the main text: oscillations get significantly suppressed at low in-plane B field ($\sim 0.7T$)

\begin{figure}[htbp]
		\includegraphics[width=8.6cm]{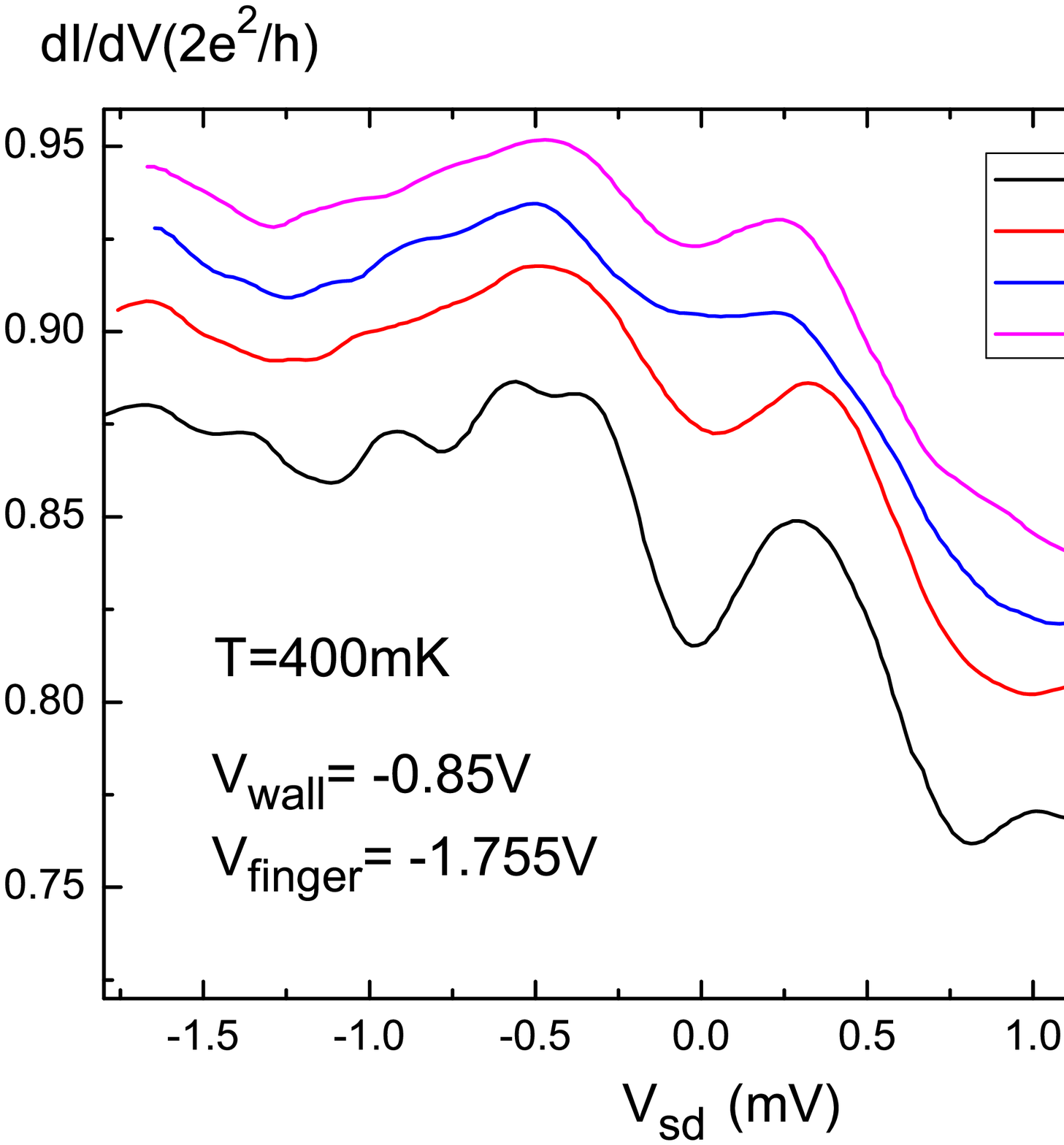}
		\caption{(Color online) The differential conductance of the 100nm channel length asymmetric QPC, at different in-plane B fields, at T=400mK}
\end{figure}